\documentstyle[12pt]{article}
\begin{document}

\thispagestyle{empty}
\begin{flushright}                              FIAN/TD/96-03\\
                                                hep-th/9602151\\
                                                February 1996

\vspace{2cm}
\end{flushright}
\begin{center}
\normalsize
{\Large\bf Remark to the  Comment on

"New pseudoclassical model for Weyl particles"}

\vspace{3ex}
{\Large D. M. Gitman$^{\dagger}$, A.E. Gon\c calves }

{\large Instituto de F\'{\i}sica, Universidade de S\~ao Paulo

P.O. Box 66318, 05389-970 S\~ao Paulo, SP, Brazil}

{\Large I. V. Tyutin$^{\ddagger}$}

{\large Department of Theoretical Physics, P.~N.~Lebedev Physical
Institute,

  Leninsky prospect 53, 117924 Moscow, Russia}

\vspace{5ex}

\end{center}

\centerline{{\Large\bf Abstract}}


\begin{quote}

We present here our considerations concerning the
problem of classical consistency of pseudoclassical models
touched upon in a recent comment
on our paper "New pseudoclassical model for Weyl particle".

\end{quote}

\vfill
\noindent

$^\dagger$ E-mail address: gitman@snfma1.if.usp.br

$^\ddagger$ E-mail address: tyutin@lpi.ac.ru

\newpage


\baselineskip=6mm

\setcounter{page}{1}

In a recent paper \cite{b1} we have proposed a pseudoclassical
action to describe massless spin one-half particle in 3+1
dimensions. Quantization of this action reproduces the minimal
theory of Weyl particle. The action has the form
\begin{equation}\label{e1}
S=\int\limits_0^1\left[-\frac{1}{2e}
\left(\dot x^\mu-i\psi^\mu\chi+i\varepsilon^{\mu\nu\rho\zeta}
b_\nu\psi_\rho\psi_\zeta+{\alpha\over2}b^\mu\right)^2-
i\psi_\mu\dot\psi^\mu\right]d\tau\;,
\end{equation}
where $x ,\; e, b $, are even and $\psi ,\; \chi$ odd
(Grassmannian) variables, whereas $\alpha$ is an even constant.

In a comment on  our work \cite{b2} it was remarked that
the classical theory, which corresponds to the action
(\ref{e1}), is inconsistent if one regards $\alpha $ as a real
number. In spite of the fact that we did not consider $\alpha$
in that sense at the classical level we
have to assign only two real possible values $\alpha=\pm1$ to it
at the quantum level. It is likely that it is this circumstance
that was interpreted by the authors of the Comment as a
problem or, as they claim, as an inconsistency of the model at
the classical level. One ought to say that it is not for the
first time that such a problem appears in
the pseudoclassical mechanics (see, for example, \cite{Hov}) and
is certainly familiar to us.  There exist different points of
view on this problem. One of them  was advocated before by
the authors of the Comment \cite{b3} and implies, in
fact, that some classical constants (in the case under
consideration $\alpha$) should be replaced by "dynamical
variables".  In the Comment the authors offer to modify our
model \cite{b1} and the subsequent analogous models \cite{b4} in
the same manner. Such a modification is possible indeed and we
knew this way of action from the above mentioned publications.
However, our point of view is that such a modification is
not necessary, and moreover may be not relevant.  The motivation
is the following. It is known from numerous examples, very
often some restrictions for possible values of parameters arise
in the quantum theory. Such restrictions can in turn depend on
details of quantization (in particular, in \cite{b2},
$\alpha=\pm 1$, or $\alpha=\pm 2$, depending on the operators
ordering). Thus, the quantum dynamics may impose restrictions on
possible parameter values within the anticipated domain of
their variation. We believe that one has to admit the same
situation for the classical theory, and, moreover, to admit an
alteration of the parameters nature in course of transition from
the classical to the quantum theory (we do admit such an
alteration for dynamical variables). In the model (\ref{e1}), at
the classical level, $\alpha$ has  to be an even constant, of
bifermionic type. In course of quantization it becomes
into a real constant, whose possible values are defined by the
quantum dynamics.  Probably, such a treatment of the parameters
in the pseudoclassical theory could be grounded if one might
give a clear meaning to the pseudoclassical theory by
constructing a "pseudoclassical" limit of the quantum theory.

According to another point of view we write the pseudoclassical
action for purposes of the quantization only, then $\alpha$ can
be regarded from the beginning as a real number. That was done
in our recent papers \cite{b4}.

Also, following the ideas of Schwinger, we can probably
consider from the very beginning all the variables entering in
the action (\ref{e1}) as operators, and construct  their
commutation relations by means of the Schwinger--type action
principle.

In any case, the point of view according to which the nature of
parameters in pseudoclassical and quantum theory can be
different, is until now  acceptable and probably relevant to
a "pseudoclassical" limit. In this respect the possible
modifications proposed in the Comment have to be motivated by
more essential reasons.


\begin{thebibliography}{9}
\bibitem{b1}D.M. Gitman, A.E.
Gon\c calves and I.V. Tyutin, Phys. Rev. D{\bf 50}, 5439 (1994)
\bibitem{b2}J.L. Cort\'{e}s and M.S. Plyushchay, DFTUZ/96/08;
hep-th/9602106
\bibitem{Hov}P. Hove, S. Penati, M. Pernici, and P. Townsend,
Class. Quantum Grav. {\bf 6}, 1125 (1989)
\bibitem{b3}J.L.
Cort\'{e}s, M.S.  Plyushchay and L.  Vel\'{a}zquez, Phys.  Lett.
B {\bf 306}, 34 (1993)
\bibitem{b4}D.M. Gitman, A.E. Gon\c
calves and I.V.  Tyutin, Int.J.Mod.Phys.A {\bf 10}, 701 (1995);
hep-th/9601065; D.M.  Gitman, A.E. Gon\c calves,
hep-th/95106010; G.V.  Grigorian, R.P.  Grigorian and I.V.
Tyutin, hep-th/9510002; D.M.  Gitman and I.V.  Tyutin,
hep-th/9602048

\end{thebibliography}
\end{document}